\documentclass[10pt,conference]{IEEEtran}
\IEEEoverridecommandlockouts
\usepackage{cite}
\usepackage{amsmath,amssymb,amsfonts}
\usepackage{algorithmic}
\usepackage{graphicx}
\usepackage{enumitem}
\usepackage{textcomp}
\usepackage{xcolor}
\usepackage{makecell}
\usepackage{caption}
\usepackage{graphicx}
\usepackage{float} 
\usepackage[subrefformat=parens]{subfig}

\begin{document}
\bstctlcite{IEEEexample:BSTcontrol}
\title{Spatial Perceptual Quality Aware Adaptive Volumetric Video Streaming}

\author{\IEEEauthorblockN{Xi~Wang,~Wei~Liu,~Huitong~Liu~and~Peng~Yang}
\IEEEauthorblockA{Hubei Key Lab of Smart Internet, School of EIC, Huazhong University of Science and Technology, Wuhan, China \\
Emails: \{twx, liuwei, lht, yangpeng\}@hust.edu.cn}
}

\maketitle

\begin{abstract}

Volumetric video offers a highly immersive viewing experience, but poses challenges in ensuring quality of experience (QoE) due to its high bandwidth requirements. In this paper, we explore the effect of viewing distance introduced by six degrees of freedom (6DoF) spatial navigation on user's perceived quality. By considering human visual resolution limitations, we propose a visual acuity model that describes the relationship between the virtual viewing distance and the tolerable boundary point cloud density. The proposed model satisfies spatial visual requirements during 6DoF exploration. Additionally, it dynamically adjusts quality levels to balance perceptual quality and bandwidth consumption. Furthermore, we present a QoE model to represent user's perceived quality at different viewing distances precisely. Extensive experimental results demonstrate that, the proposed scheme can effectively improve the overall average QoE by up to $26\%$ over real networks and user traces, compared to existing baselines.

\end{abstract}

\section{Introduction}
Recent years have witnessed a booming trend of immersive media\cite{Han2019Survey_mobile_immersive, Clemm2020Survey_immersive_comm}. Serving as specific content presentations for augmented reality, virtual reality, and mixed reality, immersive media have created a huge wave of innovation in video content and viewing experience, generating a great deal of interest. Following advancements of head-mounted display (HMD) and three-dimensional (3D) capture hardware, volumetric video came into the picture, allowing users to move freely, not only in terms of head movement but also traveling inside the scenes. 

\begin{figure}[htbp]
    \centering
   \includegraphics[width=0.5\textwidth]{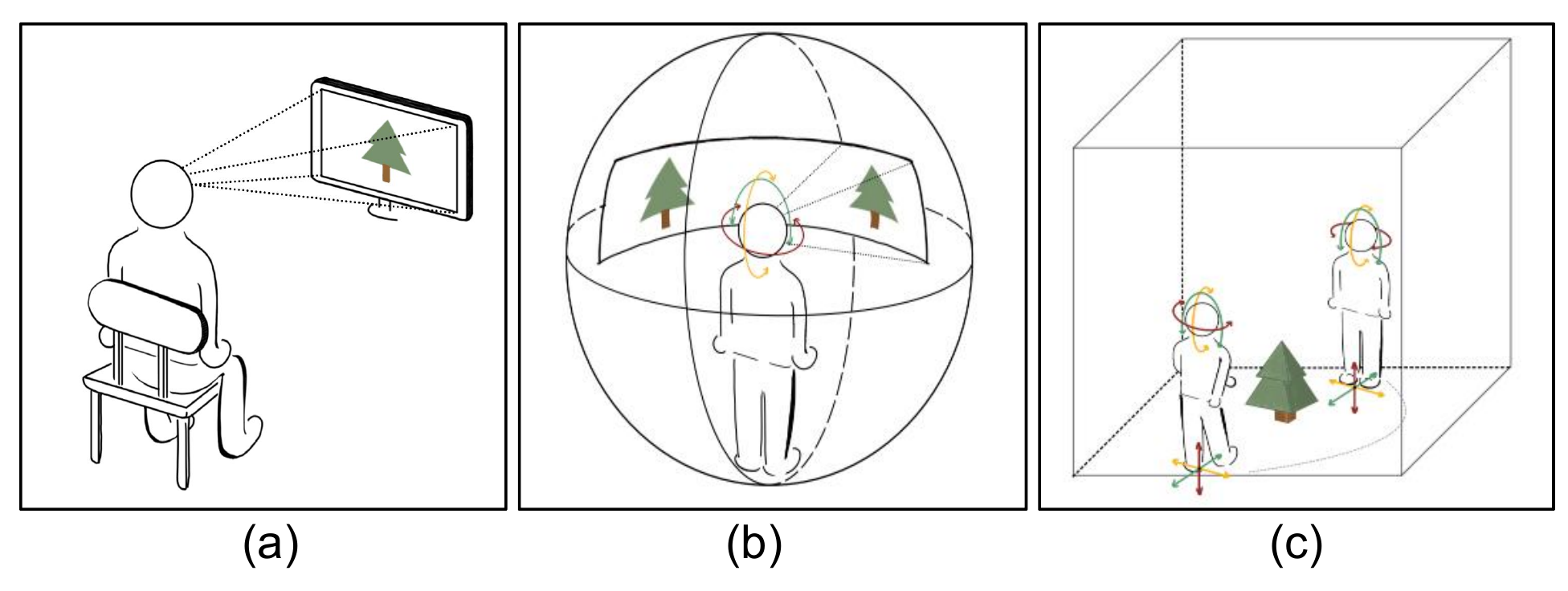}
    \caption{Comparison of user experience: (a) $2$D video, (b) $360^{\circ}$ video, (c) volumetric video.}\label{difference}
\end{figure}

Different from $360^{\circ}$ video and regular 2D video, volumetric video has two key characteristics. Firstly, volumetric video has a huge amount of data due to its unique 3D data format. At present, the most prevalent data format of volumetric video is the point cloud, attributed to its flexibility and simplicity\cite{Liu2021Survey_pointcloud_streaming}. Secondly, volumetric video can provide users with a six degrees of freedom (6DoF) experience, as shown in Fig. \ref{difference}. The users are free to change position and orientation while watching the volumetric video. The user's location and orientation relative to the viewed content determine whether the content is visible, what can be seen and whether it can be seen clearly.

%


To mitigate the bandwidth consumption brought by 3D content, some scholars\cite{park2018volumetric,Bo2020Vivo, subramanyam2020user} propose solutions to transmit only content within the visible viewing area. The generic delivery process is that volumetric video content is segmented into different groups of frames (GoFs) along the timeline, and each GoF is spatially divided into tiles, which can be compressed separately at different bitrates\cite{park2018volumetric}. The tiles within the viewport are delivered with higher quality\cite{Li2023Towards}, while the tiles outside the viewport are delivered at lower quality or discarded\cite{li2022optimal}. As an important milestone, Bo \emph{et al.}\cite{Bo2020Vivo} present the first practical mobile streaming system for volumetric video, and implement three efficient visibility optimization algorithms. However, with the inclusion of additional 3DoF information, how to continuously and dynamically adjust the strategy to accommodate the user's 6DoF behavior remains a challenging issue.

\begin{figure*}[htbp]
    \centering
    \subfloat[]{\label{dynamic_viewport}\includegraphics[width=0.45\textwidth]{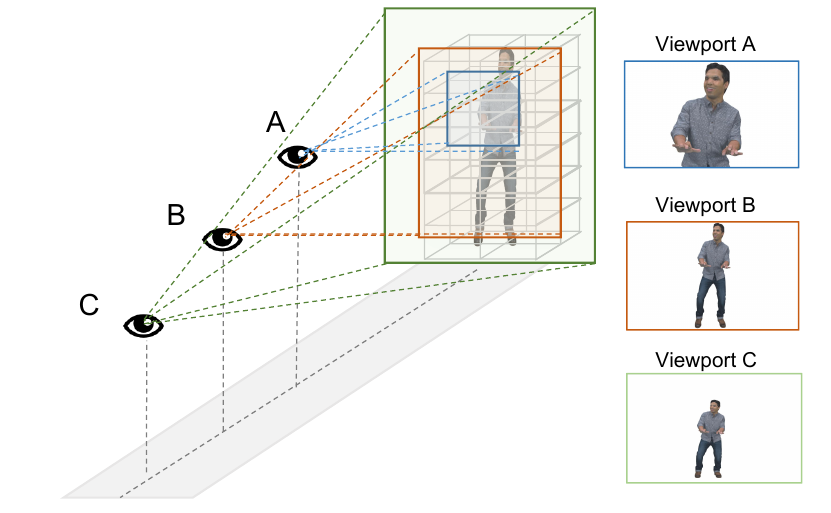}}
    \hfill
    \subfloat[]{\label{fig:system}\includegraphics[width=0.5\textwidth]{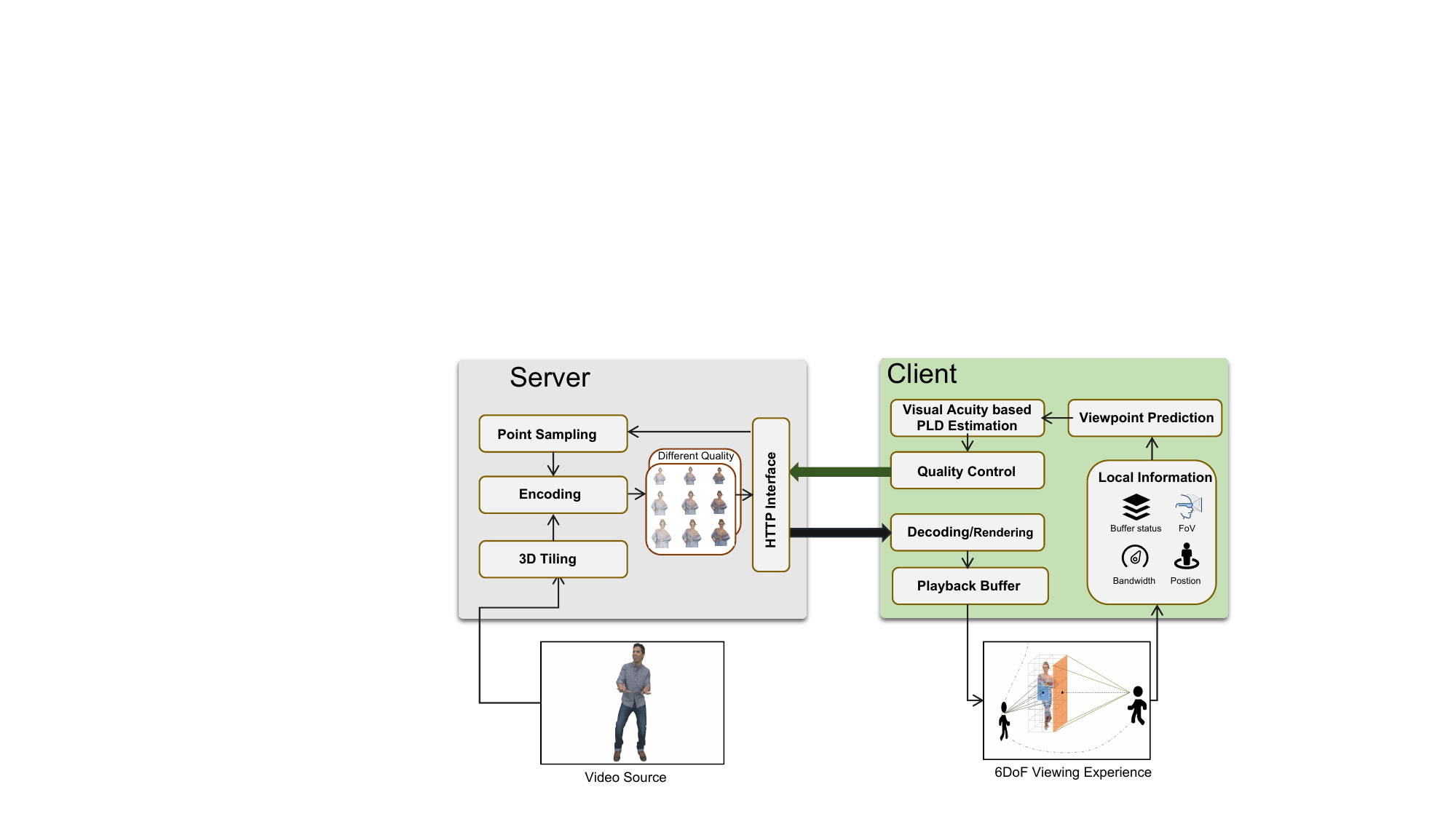}}
    
     \caption{Illustration of adaptive volumetric video streaming: (a) dynamic viewport at different viewing distances, (b) visual acuity-based adaptive streaming scheme.}
\end{figure*}

This paper aims to enhance the user's quality of experience (QoE) by leveraging a simple observation: As shown in Fig. \subref*{dynamic_viewport}, when the user's viewing position is in close proximity to the content being viewed, the viewport contains less video content and more detail can be seen. Immersive experiences make users more sensitive to visual quality\cite{yang2023delay}. To quantify the variations in perceived quality resulting from the change in viewing distance, we introduce the visual acuity of the human visual system. Consequently, further understanding of how the perceived quality of the user in space varies with viewing distance change is required. 

In this paper, we investigate the dynamic changes in the perceptual quality caused by the user's spatial exploration behavior. To elucidate this phenomenon, we formulate the visual acuity model based on the voxel size that users can distinguish at different distances. By extending the proposed model to the typical volumetric video streaming scheme, we propose a novel scheme to further enhance the user's QoE. In summary, the main contribution of this work is as follows:
\begin{itemize}
\item We propose a novel visual acuity model, which describes the relationship between viewing distance and the distinguishable boundary point cloud density (PLD).
\item We enhance the existing adaptive volumetric video streaming system by introducing the proposed visual acuity model.  
\item We present a new QoE model for volumetric video that systematically incorporates the extra 3DoF information from the user's viewing distance variations.
\item We evaluate the proposed method using real user navigation traces and bandwidth traces, and the experiment results show that the proposed scheme can effectively reduce rebuffering time while maintaining the spatial perceptual quality.
\end{itemize}


\section{Related Works}\label{2}

In this section, we focus on studies related to the unique visual visibility of volumetric video.

\textbf{Visibility optimizations.} How to effectively utilize the viewing properties of volumetric video in the process of adaptive transmission has drawn increasing attention recently. Considering human visual acuity, Hosseini \emph{et al.}\cite{hosseini2018dynamic} propose sub-sampling algorithms to obtain various quality representations for point cloud. The study in Ref.\cite{park2019rate} defines the utility of video content tiles at different locations and introduces a greedy algorithm for bitrate allocation between tiles of multiple volumetric objects. In addition, Wang et al.\cite{wang2021qoe} introduce perspective projection to give a geometric computational paradigm for the view frustum in virtual space. 
\textbf{New quality-determining factors.} New quality-determining factors in regular 2D and $360^\circ$ video\cite{Guan2019Pano, yang2021just} have been well explored in recent years. Viewing distance, as a determinant of perceptual quality, has gradually attracted the attention of some scholars\cite{park2019rate,Bo2020Vivo}. Based on the characteristics of the human visual system, video content with the same bitrate will have different perceived clarity at different physical viewing distances. Specifically, the human visual system can distinguish content with a minimum angle of less than $1^{'}$ between the adjacent pixels and the human eyes\cite{fender2017heatspace}. As the user's distance from the viewing content increases, the voxel size can be dynamically adjusted to maintain a similar visual experience. While visual acuity varies individually and changes with age, the normal visual acuity of adults is $1'$ in size, or 60 pixels per degree\cite{INTRILIGATOR2001171}.


Most of the existing work selects the PLD version by fixing the viewing distance interval, which makes the quality selection sensitive to the distance parameters. Therefore, the relationship between human spatial perceptual quality at different locations, PLD and voxel size needs further investigation.

\section{Visual Acuity Based Volumetric Video Streaming System}\label{3}
In this section, we introduce spatial perceptual quality and propose an adaptive streaming scheme based on visual acuity.
\subsection{System Architecture}
We consider the adaptive volumetric video streaming system consisting of client side and server side, as shown in Fig. \subref*{fig:system}. The server-side stores the source videos at various PLDs. Meanwhile, after the client sends a request, the server side predicts the bandwidth and user's 6DoF trajectory using the Long Short-Term Memory network based on the local information (including bandwidth, user's previous trajectories) provided by the client. Based on the predicted user trajectory, we introduce a spatial perspective projection\cite{wang2021qoe} to calculate the tiles contained in the user's view frustum, and we utilize a visual acuity model to compute the distinguished boundary PLD. The server side determines on the quality level of tiles to be transmitted based on the bandwidth and calculated distinguished boundary PLD. After the compressed tiles and Media Presentation Description files are sent to the client, the client decodes and renders the tiles and stores them in a buffer.
\subsection{Visual Acuity Model}

For volumetric video streaming, a higher PLD usually means a higher bitrate and better perceptual quality. When the transmitted video resolution exceeds the limits of human visual acuity or resolution provided by the device, there will be a gap between the transmitted video quality and perceptual video quality. As the user is allowed 6DoF motion during viewing, the viewport size changes dynamically with the viewing distance, which leads to further accumulation of the difference between the transmitted and perceived video quality, resulting in wasted bandwidth and increased latency. Therefore, we propose the visual acuity model to calculate the boundary PLD that can be distinguished by the user's naked eye at the current spatial location.


When the user is watching the volumetric video, sensors such as gyroscopes in the helmet record the user's motion information, which will be used to adjust the camera position in the virtual world. The initial camera position is determined by the size of the convex closure of volumetric content in 3D space and the optimal viewing distance of the device. As illustrated in Fig.\ref{new}, the initial distance is denoted as $d_0$ (default value), with the voxel size at this distance being equivalent to the highest resolution of the original video, denoted as $v_0$. We first consider the maximum planar resolution that can be provided by human visual projection at different viewing distances for the user. In this case, the user viewing distance becomes farther. The effective percentage of video content in the viewport decreases, while the detail that can be seen by the human eye becomes less. Since the minimum visual angle $\theta$ of adjacent pixels resolved by the human eye is fixed, we use the minimum Pixel per Inch (PPI) to represent human visual acuity at the initial distance $d_0$:

\begin{equation}\label{initial}
P P I_0=\frac{1}{2 \times d_0 \times \tan \left(\frac{1}{2} \times \frac{1}{60} \times \frac{\pi}{180}\right)}
\end{equation}

%
Therefore, $PPI_0$ can reflect the maximum number of pixels that people can see clearly from the plane and space perspective. The information in the viewed content is calculated based on human visual acuity and display resolution\cite{fender2017heatspace}. Suppose $d_t$ is the distance from the camera position to the centre of volumetric content object during the playback of segment $t$. Then, the minimum PPI of segment $t$ can be given by:
\begin{figure}[t]
 \centering
 \includegraphics[width=0.5\textwidth]{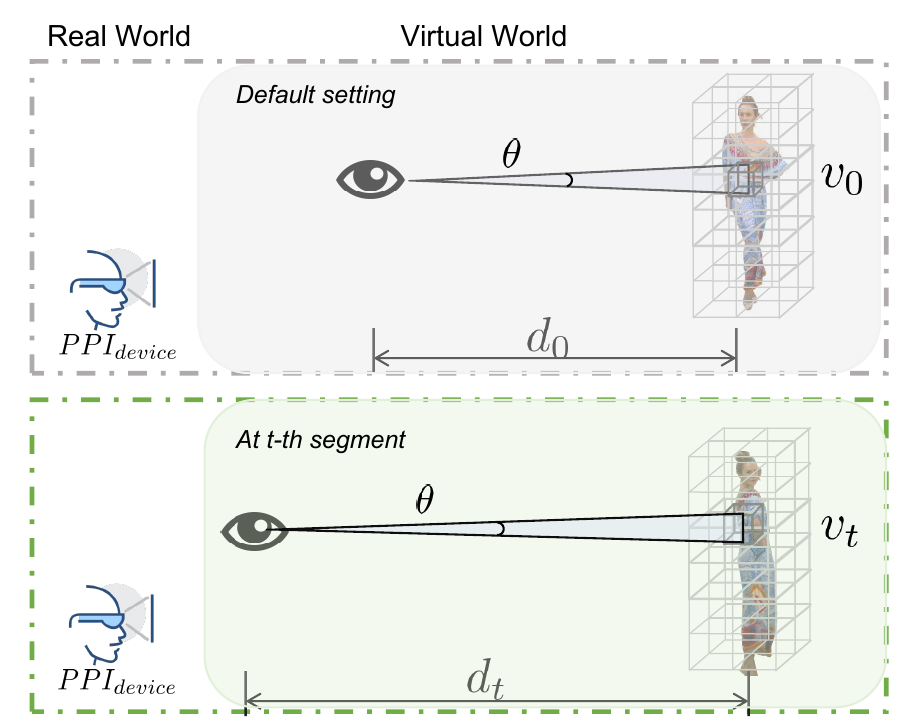}
 \caption{The variation between distinguishable voxels when the distance is different in virtual world.}
 \label{new}
\end{figure}
\begin{equation}\label{initial2}
P P I_t=\frac{d_0}{ d_t}\times PPI_0 
\end{equation}

From Eq.(\ref{initial2}), we can obtain that the number of pixels seen by the user decreases as the distance increases. It is worth noting that the number of pixels seen by the user does not keep getting larger as the user gets closer. Here, we introduce a discussion of the maximum resolution of the device, the minimum $PPI_t ( d_t < d_0)$ can be expressed as:
\begin{equation}\label{initial3}
P_t=\min (PPI_t, PPI_{device}) 
\end{equation}
where $PPI_{device}$ is a fixed parameter determined by the device hardware. Based on the change in the number of pixels per inch, the distinguished voxel size $v_t$ at segment $t$ can be further expressed as:
\begin{equation}\label{initial3}
v_t \cdot P_t \doteq {P P I_0} \cdot{v_0}
\end{equation}

Next, we use octree to voxelize the point cloud with specified voxel sizes, and we can use a function  $\mathcal{H}(\cdot)$ to denote the mapping relationship between PLD and voxel size. The boundary PLD at segment $t$ can be expressed as:
\begin{equation}\label{initial3}
\eta^{*}_t = \mathcal{H}^{-1}(\frac{P P I_0 \cdot v_0}{P_t})
\end{equation}

$\eta^{*}_t$ is used as a boundary value to guide the quality level selection. When the requested PLD is less than $\eta^{*}_t$, the user can perceive a decrease in spatial resolution. Correspondingly, the user's movement in space can make the change in viewing video content ignored.

\section{Problem Formulation}

For adaptive video transmission, a generic QoE model consists of perceived video quality, rebuffering time, inter- and intra-segment quality variations\cite{qian2018flare,feng2022perceptual}. In the proposed system, we synergistically integrate the visual acuity model with the general QoE model to enhance the four factors: perceived video quality, rebuffering time, spatial quality variation and temporal quality variation. This approach introduces spatial perceptual quality, providing a more detailed and accurate evaluation of the QoE under 6DoF motion.

In the proposed system, the source video is segmented into $T$ GoFs in temporal domain, and each GoF consists of $L \times W \times H$ tiles in spatial domain. Let $\mathcal{T} = \{t | t\in \left( 1,2, \ldots, T\right)\}$ denote the set of GoFs. Let $\mathcal{I} = \{i | i\in \left( 1,2, \ldots, L \times W \times H\right)\}$ denote the set of tiles of each GoF. Let $m_{t,i}$ denote the sampling factor for tile $i$ of GoF $t$, which represents the ratio of downsampling. Then, each tile $i \in\{1,2, \ldots, L \times W \times H\}$ is independently upsampled to different quality level at different coefficient $m_{t,i}$. Let an indicator matrix $v_{t, i} \in \{0, 1\}^{K \times L \times W \times H}$ represent whether tile $i$ of GOF $t$ is inside the view frustum. If $v_{t, i} = 1$, the tile is inside the view frustum; otherwise, $v_{t, i} = 0$.  
\textbf{Perceived video quality.} We consider the following factors using a weighted V-PSNR to represent the perceptual video quality. Firstly, we only consider the effect of tiles in the view frustum on the perceived video quality\cite{li2022optimal}. Secondly, we consider tiles whose PLD meets the visual acuity constraint to have a positive impact on the video quality. It is noted that we obtain the PSNR lookup table by collecting a large number of snapshots of the real rendering results for different point cloud densities at different viewing distances. The discrete values of the PSNR lookup table are fitted to a continuous curve so that we can obtain the PSNR values for any discrete viewing distance. The perceived video quality of segment $t$ can be calculated as follows:
\begin{equation}
Q^{1}_{t}= \frac{\sum_{i=1}^{L\times W \times H}  v_{t,i} \cdot  PSNR(m_{t,i}, d_{t,i}) \cdot \mathbb{I}_{\left\{\eta_n \geq \eta^{*}_t)\right\}} } {\sum_{i=1}^{L\times W \times H}  v_{t,i} }
\end{equation}
where $\mathbb{I}_{\{.\}}$ represents whether the quality change can be ignored when the sampling factor is $m_{t,i}$. Specifically:

\begin{equation}
\mathbb{I}_{\{.\}}=\left\{\begin{array}{lr}
1,                                  &\text{if }\{.\}\text{is true } \\
\frac{\eta(m_{t,i})}{\eta^{*}_t}, &\text{otherwise }
\end{array}\right.
\end{equation}

\textbf{Rebuffering Time}: The huge data amount of volumetric video will bring non-negligible transmission delay, and the accumulation of transmission delay will lead to freezing on the client side, which has a great negative impact on user experience. The transmission time of GoF $i$ can be calculated as:
\begin{equation}
	\tau_{t} = \sum_{i=1}^{L\times W\times H}\frac{v_{t,i}\times \mathbb{S}(m_{t,i})}{BW_t}
\end{equation}
where $BW_t$ is the average bandwidth when downloading GoF $t$, and $\mathbb{S}(m_{t,i}) $ is a function reflecting the mapping between the sampling factor $m_{t,i}$ and the size of tile $i$ of GoF $t$\footnotetext[1]{Given that the size of the point cloud file exhibits a linear relationship with the number of points\cite{Lee2020GROOT_fidelity_streaming}, it is possible to calculate the tile's size at different sampling factors by multiplying the sampling factor with the size of the original file}. Let $B_t$ denote the buffer occupancy when the $t$-th GoF starts to be sent. Thus, the rebuffering time for $t$-th GoF is calculated as follows:


%
\begin{equation}
	Q_{t}^{2} = \max\{\tau_t - B_t,0\}
\end{equation}

\textbf{Temporal Quality Variation}: Since the user's transient 6DoF motion during volumetric video viewing causes large variations in the tiles being viewed, we consider the variation in the quality of tiles within the view frustum between consecutive GoFs:
\begin{equation}
	Q_{t}^{3} = \left| Q_{t}^{1} - Q_{t-1}^{1} \right|
\end{equation}

\textbf{Spatial Quality Variation}: For the interior of a chunk, the quality difference between tiles within the view frustum also affects the user's viewing experience. The quality variation of spatial tiles in the same chunk can be expressed by the following:
\begin{equation}
Q_{t}^{4} =Std \{PSNR(m_{t,i}, d_{t,i})  \cdot \mathbb{I}_{\left\{\eta_n \geq \eta^{*}_t\right\}} \mid \forall i: v_{t,i} = 1\}
\end{equation}



Based on these factors, the volumetric video streaming system can be formulated as a PLD selection problem between the server and the client:


\begin{equation}\label{final}
\mathrm{P}_0:  \max QoE = \sum_{t=1}^{K}  Q_{t}^{1} - p Q_{t}^{2} - q Q_{t}^{3} - r Q_{t}^{4} \\
\end{equation}

\begin{equation}\label{constraints1}
\text { s.t. } \sum_{i} \mathbb{S}(m_{t,i}) \leq \sum_{i}  S_{t,i} , \forall  t \in \mathcal{T} \\
\end{equation}

\begin{equation}
\text  m_{t,i} > 0,  v_{t, i}\in\{0,1\}, \\
\end{equation}

The overall objective is the weighted sum of the three described QoE factors, where $p$, $q$ and $r$ are the weight parameters reflecting the preference. For the proposed optimization problem, we use Eq.(\ref{constraints1}) to describe the transmission constraint, which means the actual size of the transmitted tiles must be less than the total size of this chunk. The proposed QoE optimization problem can be considered as a knapsack problem that can be solved by dynamic programming and heuristic methods. It remains a challenge due to the large search space. In order to improve the real-time performance of the algorithm, we need to prune the search space efficiently. We use the boundary PLD derived from the visual acuity model to prune the search space and a greedy-based method to solve the problem.

%

\section{Performance Evaluation}\label{5}
In this section, we perform experiments with user trajectory-based simulations to verify the effectiveness of the proposed scheme.

\subsection{Experiment Setting}
\textbf{Dataset}: In our experiments, we use an existing 6DoF user navigation trajectory dataset\cite{	Liu2021Survey_pointcloud_streaming}, which includes trajectories of $40$ users watching $4$ point cloud videos captured at $30$ frames per second. The chunk duration is approximately $333$ ms, and each video chunk is partitioned into $64$ equal-sized tiles. The initial viewing distance is set to $d_0 = 1m$, while the initial FoV spans $110^{\circ}\times 110^{\circ}$. To obtain versions with different PDLs, we used the Draco library\cite{draco2023} for compressing at bitrates in $\{87.5, 161.6, 296.4, 420.7, 565.2, 651\}$ Mbps (which pretains to the PLD at $\{10\%, 20\%, 40\%, 60\%, 80\%, 100\% \}$). Real-world network traces are randomly selected from the 5G dataset\cite{raca2020beyond} to evaluate the performance of the proposed system.

\textbf{Parameters}: We set the hyper-parameters as $p=50$ and $q=r=1$ in Eq.(\ref{final}) for the QoE definition. These values indicate that a penalty of $1$s for rebuffering is considered equivalent to the average PSNR of the highest-quality video chunk played smoothly. 
\begin{figure}[htbp]
    \centering
    \subfloat[]{\includegraphics[width=0.13\textwidth,height = 0.20\textheight]{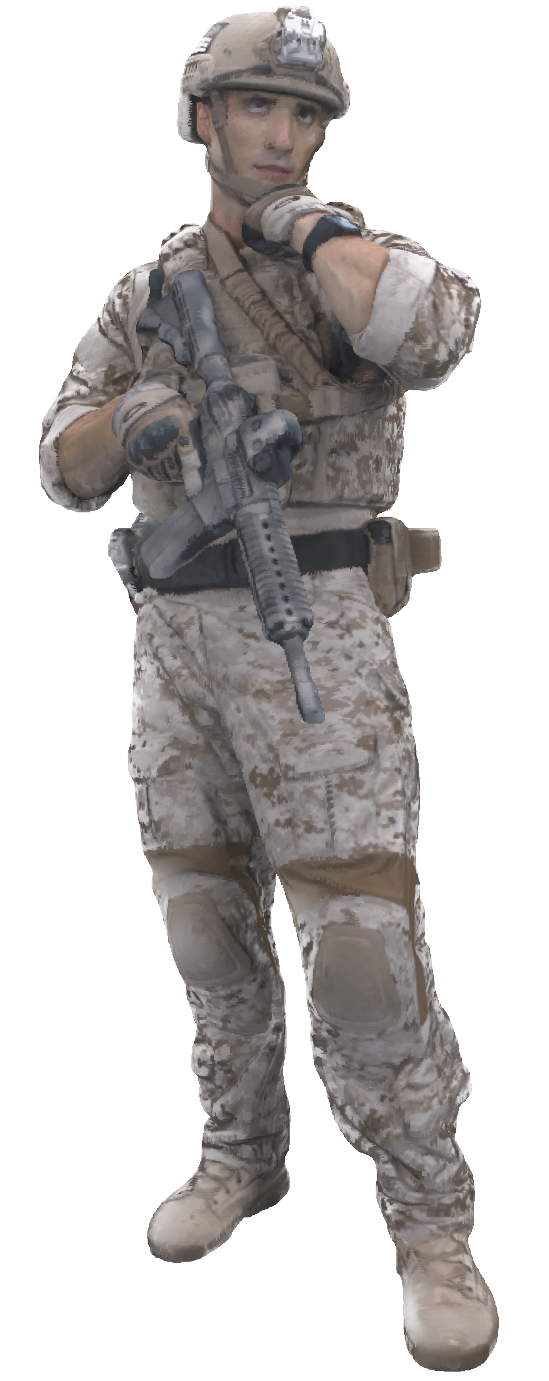}}
    \subfloat[]{\includegraphics[width=0.12\textwidth,height = 0.20\textheight]{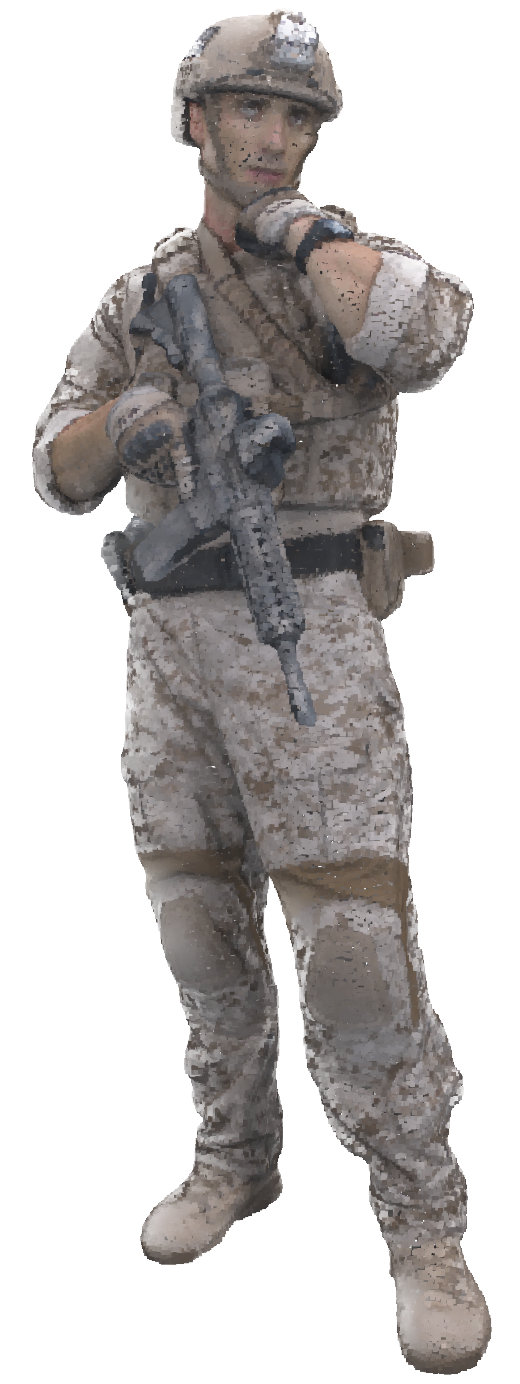}}
    \subfloat[]{\includegraphics[width=0.08\textwidth, height = 0.17\textheight]{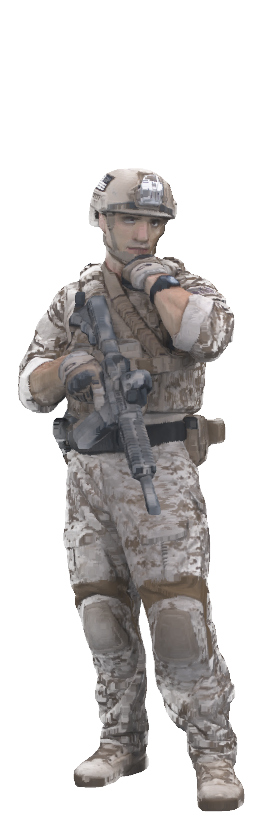}}
    \subfloat[]{\includegraphics[width=0.08\textwidth, height = 0.16\textheight]{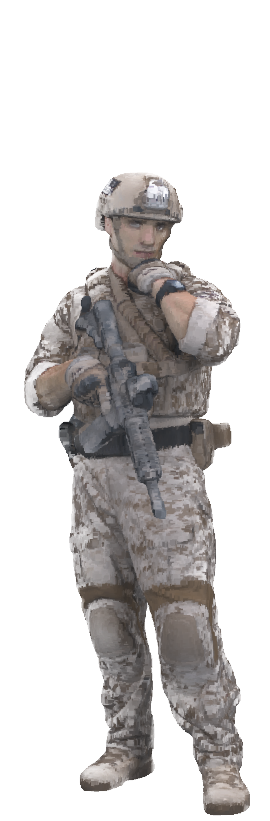}}
    \caption{Visual comparison of different quality of a specific frame at different viewing distances: (a) original point cloud (0.6m), (b) upsampled 20\% point cloud (0.6m), (c) original point cloud (1.8m), (d) upsampled 20\% point cloud (1.8m)}\label{visual}
\end{figure}

\textbf{Baselines}: We compare the proposed algorithm with the following benchmark schemes:
\begin{itemize}
	\item \emph{rate \& utility}: Based on a greedy algorithm, the bits of different tiles are allocated according to the utility defined by the relationship and distance between the tiles and the user's view frustum\cite{park2019rate}.
	
	\item \emph{viewport \& utility}: A QoE-driven and tile-based adaptive point cloud streaming\cite{wang2021qoe}, considering the view frustum calculated by perspective projection.

    \item \emph{distance \& tile}: Based on the discrete mapping of the distance range to the PLD, each tile in the view frustum is assigned a quality level, similar to the distance visibility algorithm proposed in \cite{Bo2020Vivo}.
\end{itemize}

%
%
\subsection{Experimental Results}

\subsubsection{Visual Effects Comparison} Fig. \ref{visual} presents two groups of snapshots from our experiments using fixed angles at varying distances on video "Soldier" with different PLDs. It illustrates that, even when the original point cloud video frame is downsampled at $20\%$, changes in the user's viewing distance can lead to significant quality effects that are negligible or even unnoticeable.

\subsubsection{Individual Case Studies} 
\begin{figure}[htbp]
    \centering
  
    \subfloat[]{\includegraphics[width=0.25\textwidth, height = 0.16\textheight]{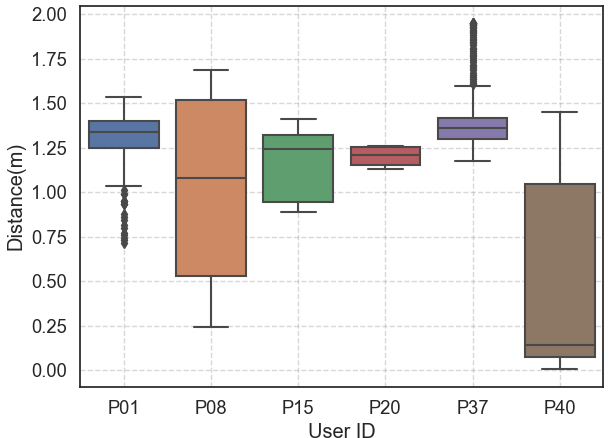}}
    \subfloat[]{\includegraphics[width=0.25\textwidth, height = 0.16\textheight]{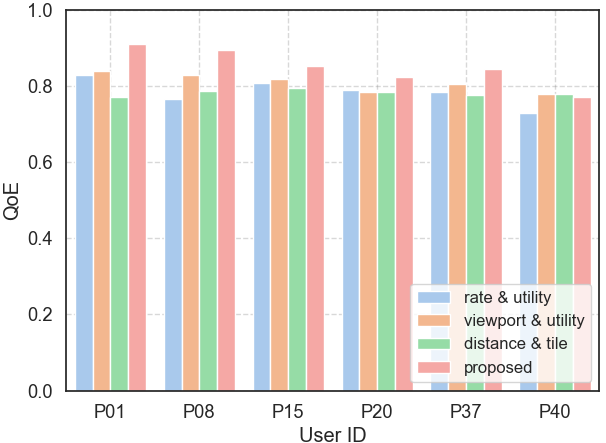}}
    \caption{Comparison of user case quality between different schemes: (a) users' viewing distance, (b) average QoE (nomarlized) among different users.}\label{distance_qoe}
\end{figure}

To perform quantitative analysis, we randomly selected 6 users' viewing trajectories for comparative experiments. As shown in Fig. \ref{distance_qoe}, all users with different moving trends almost have some performance improvement. It is worth stating that we intentionally find a case where the proposed scheme performs the worst, the viewing distance user P$40$ is usually positioned at a significantly shorter distance compared to the initial viewing distance $d_0$. Therefore, the proposed method sacrifices some rebuffering time to maintain the perceptual visual quality, which leads to a slight decrease in the normalized QoE.

\begin{figure}[htbp]
    \centering
    \subfloat[]{\label{datasets1}\includegraphics[width=0.24\textwidth, height = 0.16\textheight]{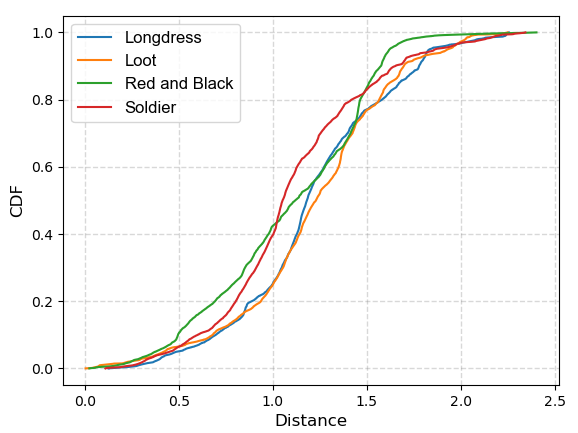}}
    \subfloat[]{\label{datasets2}\includegraphics[width=0.24\textwidth, height = 0.16\textheight]{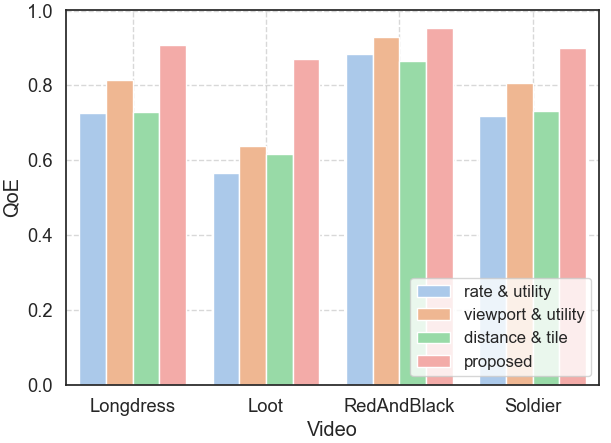}}

    \caption{Performance comparison among different videos: (a) CDFs of viewing distance,(b) average QoE among different video.}\label{datasets}
\end{figure}

\subsubsection{Performance Comparison of Typical Videos} To validate the performance of different content types, we conduct comprehensive experiments using various video datasets. Fig. \subref*{datasets1} shows the distribution of distances at which users watch different video contents. As illustrated in Fig. \subref*{datasets2}, the proposed scheme outperforms three baselines on different video contents. Especially, for the video "Loot", the distribution probability of the user viewing distance $d > 1.35$ is the highest, and the proposed scheme improves the QoE more significantly. The proposed method consistently yields QoE improvements ranging from $6\%$ to $26\%$ compared to three baselines across different video contents.

\begin{figure*}[htbp]
    \centering

    \subfloat[]{\label{qoe-2}\includegraphics[width=0.25\textwidth, height = 0.16\textheight]{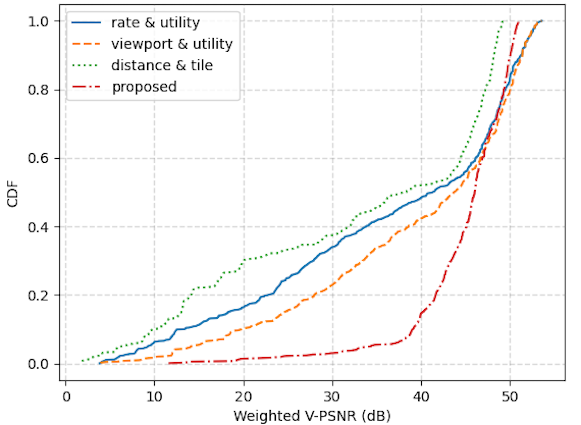}}
    \hfill
    \subfloat[]{\label{qoe-3}\includegraphics[width=0.25\textwidth, height = 0.16\textheight]{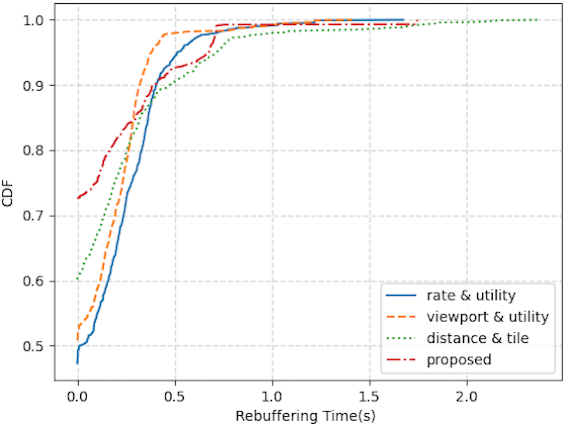}}
     \hfill
    \subfloat[]{\label{qoe-4}\includegraphics[width=0.25\textwidth, height = 0.16\textheight]{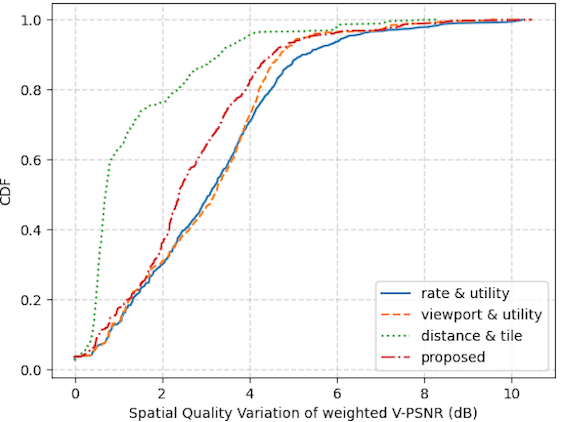}}
     \subfloat[]{\label{qoe-5}\includegraphics[width=0.25\textwidth, height = 0.16\textheight]{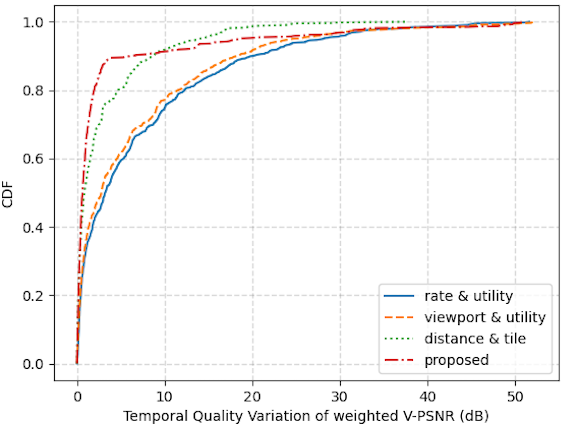}}
     \caption{Performance comparison among different schemes: (a) weighted viewport PSNR, (b) rebuffering time, (c) spatial variation, (d) temporal variation.}\label{qoe}
\end{figure*}
\subsubsection{Overall Analysis among QoE Factors} To better understand the QoE gains for the overall users, we analyze the performance of all individual QoE factors. Fig. \ref{qoe} presents a comparison among different schemes of video quality, rebuffering time, video temporal and spatial quality variance. It is worth stating that a small fraction of users will have a weaker V-PSNR than \emph{rate \& utility} and \emph{viewport \& utility}, which is due to the mechanism of assigning quality. As shown in the Fig. \subref*{qoe-2}, the divergence occurs when the weighted V-PSNR exceeds $49$ dB. This discrepancy can be attributed to the fact that the proposed scheme is customized to the user behavior rather than solely focusing on the available bandwidth. Therefore, with adequate bandwidth, \emph{rate \& utility} and \emph{viewport \& utility} will prioritize transmitting higher quality tiles, while the proposed scheme will cease enhancing quality after meeting the user's perceived quality requirements. Fig. \subref*{qoe-3} shows the rebuffering time among different schemes. It can be noted that the proposed scheme will be close to the \emph{distance \& tile} in a few cases, that the higher quality tiles will be requested when the user is walking near to viewed content. Fig. \subref*{qoe-4} and \subref*{qoe-5} show the spatial and temporal quality variation of tiles within the viewport, respectively. Since the \emph{distance \& tile} assigns quality levels based on the distance between the user and each tile, it achieves better results in terms of quality variation as the distance varies smoothly.

\section{Conclusion}\label{6}
In this paper, we have explored the user's 6DoF spatial navigation behavior and proposed a visual acuity model to calculate tolerable boundary PLD under varying distances. Leveraging the 6DoF information, we have adaptively determined the appropriate video content and quality level for transmission. By incorporating the proposed model into an adaptive volumetric video streaming system, we have proposed an effective QoE model that accounts for viewing distance variation. Our evaluation has demonstrated that, the proposed scheme improved the average weighted V-PSNR by $7 -12$ dB and effectively reduced the rebuffering time under the real-world 5G bandwidth trajectory, in comparison to the existing solutions. While personalized QoE may be influenced by individual viewing preferences and video contents, the proposed scheme significantly has improved the average QoE by up to $26\%$ across various video contents. In the future, we plan to investigate the visibility optimization in full scene volumetric video.

%

\bibliographystyle{IEEEtran}
\bibliography{main}

\end{document}